\documentclass{raa}           
\usepackage{graphicx,times}
\usepackage{natbib}
\usepackage{amssymb,amsmath}
\bibpunct{(}{)}{;}{a}{}{,}
\usepackage{commath}
\usepackage[pagebackref=true]{hyperref}
\hypersetup{pdftitle = CALLISTO-Peru, pdfauthor = J Rengifo, pdfsubject= Radio observations, pdfkeywords = CALLISTO SRB RFI} 
\hypersetup{colorlinks = true, linkcolor = green, anchorcolor = red, citecolor = blue, filecolor = red, urlcolor = red}

\begin{document}

\title{CALLISTO Facilities in Peru: Spectrometers Commissioning and Observations of Type III Solar Radio Bursts}

\volnopage{ {\bf 202X} Vol.\ {\bf X} No. {\bf XX}, 000--000}
\setcounter{page}{1}

\author{J.A. Rengifo\inst{1,2}, V. Loaiza-Tacuri\inst{3,2}, 
  J. Bazo\inst{1}, W.R. Guevara Day\inst{4,2} }

\institute{ Sección F\'isica, Departamento de 
Ciencias, Pontificia Universidad Cat\'olica del Per\'u, Av. Universitaria 1801, Lima 32, Peru; {\it jbazo@pucp.edu.pe}\\
    \and
    Direcci\'on de Astrof\'isica, Comisi\'on Nacional de Investigaci\'on y Desarrollo Aeroespacial, CONIDA, Luis Felipe Villar\'an 1069, Lima, Peru\\
	\and
    Observat\'orio Nacional, Rua General Jos\'e Cristino, 77, 20921-400 São Crist\'ovão, Rio de Janeiro, RJ, Brazil\\
    \and 
    Facultad de Ciencias F\'isicas, Universidad Nacional Mayor de San Marcos, P.O. Box 14-149, Lima 14, Peru\\
    \vs \no
   {\small Received 202X Month Day; accepted 202X Month Day}
}

\abstract{
The Astrophysics Directorate of CONIDA has installed two radio spectrometer stations belonging to the e-CALLISTO network in Lima, Peru. Given their strategic location near the Equator, it is possible to observe the Sun evenly throughout the whole year. The receiver located at Pucusana, nearby the capital city of Lima, took data from October 2014 until August 2016 in the metric and decimetric bands looking for radio bursts. During this period, this e-CALLISTO detector was unique in its time-zone coverage. To assess the suitability of the sites and the performance of the antennas we analyzed the radio ambient background and measured their radiation pattern and beam-width. To show the capabilities of the facilities to study solar dynamics in these radio frequencies we have selected and analyzed type III Solar Radio Bursts. The study of this kind of bursts helps to understand the electron beams traversing the solar corona and the solar atmospheric density. We have characterized the most common radio bursts with the following mean values: a negative drift rate of -25.8 $\pm$ 3.7 MHz/s, a duration of 2.6 $\pm$ 0.3 s and 35 MHz bandwidth in the frequency range of 114 to 174 MHz. In addition, for some events, it was possible to calculate a global frequency drift which on average was 0.4 $\pm$ 0.1 MHz/s. 
}

\authorrunning{J. Rengifo et al.}    
\titlerunning{CALLISTO Facilities in Peru}  
\maketitle


\section{Introduction}
\label{Sec_intro}

The CALLISTO (Compound Astronomical Low frequency Low cost Instrument for Spectroscopy and Transportable Observatory) (\cite{Benz2005143,Benz2009277,CALLISTOweb}) radio spectrometers make the e-CALLISTO network with more than 170 stations, while, on average, 60 provide data every day around the world. The main scientific goal of the network is to monitor all types of solar radio activities at a wide range of frequencies (i.e. from 45 to 870 MHz). 

Two CALLISTO spectrometers with an LPDA (Logarithmic Periodic Dipole Array) antenna were installed and commissioned, between 2012 and 2014, in two sites in Lima, Peru, by the Astrophysics Directorate of CONIDA (Comisión Nacional de Investigación y Desarrollo Aeroespacial). The stations were part of an agreement of the International Space Weather Initiative (ISWI), sponsored by the United Nations Office for Outer Space Affairs (UNOOSA), NASA and JAXA, at the 50$^{th}$ anniversary meeting of the International Geophysical Year (International Heliophysical Year (IHY 2007)). The receiver with lower background that we analyzed in this work took data from October 2014 until August 2016. During this period, this e-CALLISTO station in Peru was unique in its time-zone (i.e. GMT-5) coverage. Currently, there are more instruments covering this time zone and similar stations have been deployed elsewhere (e.g. \cite{Prasert2019,Zavvari2016185}) 

We have used the recorded radio data to look for type III Solar Radio Bursts (SRBs), which are common transient bursts. They last a few seconds and have a characteristic rapid drift from higher to lower frequencies over time (\cite{Reid_2014}). We have found burst candidates from the data taking period and analyzed their spectra, showing the observational capabilities of the facilities. Here we present the results of these observations and their calculated burst parameters. Other type III SRBs have also been characterized using similar e-CALLISTO stations (\cite{Ansor2019,Hamidi2019, Hamidi2016, Ali2016, Ramli2015123}). 

This work is divided as follows: first we present in Sec. \ref{Sec_SRB} a summary of the main characteristics of type III SRBs that we searched for and studied. In Sec. \ref{Sec_commissioning} we describe the commissioning of the radio spectrometer stations, which includes a study of the radio frequency interference background in both sites and the measurement of the radiation pattern and beamwidth to characterize the antennas. Then, in Sec. \ref{Sec_Obs} we present the observations of radio bursts recorded with one of the installed e-CALLISTO stations. With the filtered data we analyze the features and dynamic spectra of type III SRBs, inferring the observational burst parameters of each isolate burst and group bursts. Finally, in Sec. \ref{Sec_conclusions} we give our conclusions.

\section{Type III Solar Radio Bursts}
\label{Sec_SRB}

Type III SRBs are common radio transient emissions, which have been studied for more than 70 years. They are observed by instruments at frequencies ranging from kHz (\cite{kasahara_2001}) up to 7.6 GHz (\cite{Ma_2012}). They are characterized by a fast drift from higher to lower frequencies over time (\cite{Reid_2014}). Their frequency drift rate has been deeply analyzed (\cite{Suzuki1985}).

Type III bursts can occur singularly, in groups or in storms and can be accompanied by a second harmonic (i.e. another event produced at twice the frequency of the first one (\cite{Labrum})). The duration for single bursts is 1-3s, for bursts in groups ranges from 1 to 5 minutes and in storms from minutes to hours (\cite{Prasert2019}).

These radio emissions are produced by non-thermal electrons accelerated in the solar corona. They are considered to originate near to the particle acceleration regions during solar flares (\cite{Meszarosova2008}). Usually, low–frequency solar radio emission occurs in the solar atmosphere, including flares, coronal mass ejections and shocks, that produce particle fluxes that reach Earth. These radio bursts can be used to study the solar coronal plasma, to probe the solar atmospheric density and to indirectly measure the height where they occur (\cite{White}). 

Type III SRBs are signature from electron beam accelerated in the solar corona emitting radio waves (\cite{Reid_2014}). This is achieved via the plasma emission mechanism (\cite{Melrose_1980, Guinzburg_1958}). Type III SRBs are produced from magnetic reconnection regions and propagate along open magnetic field lines. The faster ones produce a bump-on-tail instability generating Langmuir waves at the local plasma frequency $f_p$, which can be converted into electromagnetic emission. The decay of Langmuir waves may produce radiation at the fundamental frequency and the scattered Langmuir waves produce another emission at the second harmonic ($2f_p$) (\cite{Reid_2014}).

These bursts are sometimes associated with active regions and flares. However, they can be seen at times when there is no activity at other wavelengths (e.g. no GOES-class events (\cite{Patrick})). Nevertheless, the association of flares with type III SRBs depends on the magnetic field configuration (\cite{Axisa_1974}).

In addition, SRBs can cause noise in sun-facing antennas, degrading or preventing over-the-horizon communications, and in air-to-ground signals with direct line-of-sight to the Sun (\cite{McAlester2014}). The importance of this noise is related, among other factors, to the solar incidence angle, the antenna pattern and the tracking algorithm (\cite{Xinan2018}).

\section{Commissioning of the e-CALLISTO Stations in Peru}
\label{Sec_commissioning}

\begin{figure}[ht]
	\begin{minipage}[c][1\width]{
	   0.45\textwidth}
	   \centering
	   \includegraphics[width=0.75\textwidth]{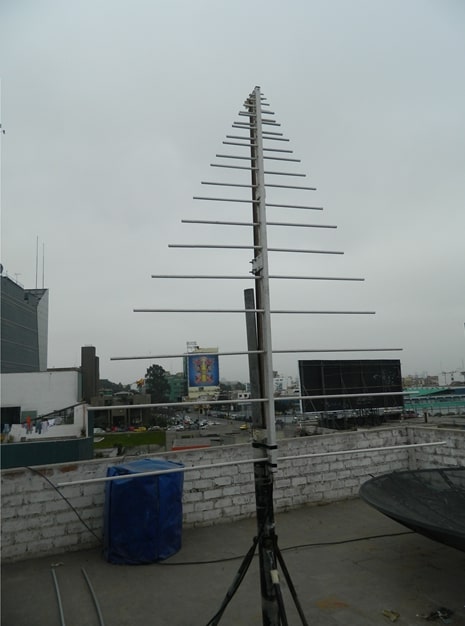}
	\end{minipage}
 \hfill 	
	\begin{minipage}[c][1\width]{
	   0.45\textwidth}
	   \centering
	   \includegraphics[width=0.75\textwidth]{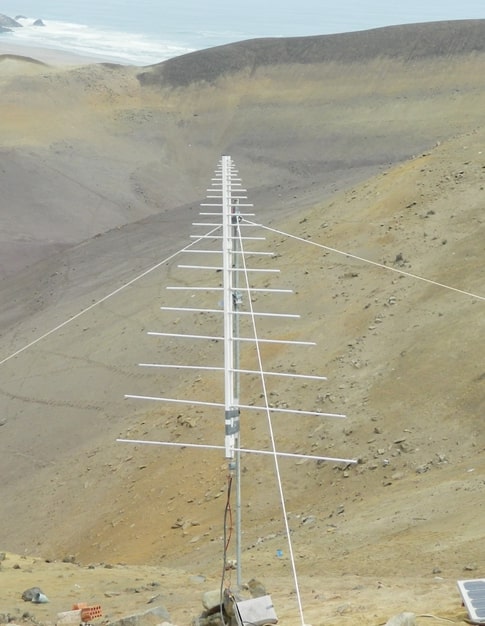}
	\end{minipage}   
	\caption{LPDA antennas installed in Lima, Peru, in city center at the San Isidro site (left) and in the city outskirts at the Pucusana site (right).}
\label{fig_antennas}
\end{figure}

Two e-CALLISTO stations, shown in Figure \ref{fig_antennas}, were commissioned in Lima, Peru, by the Astrophysics Directorate of CONIDA, as part of the global network. CALLISTO NA-06 was installed early on in 2012 in San Isidro district at the main offices of CONIDA (-12$^{\circ}$06’51” S, -77$^{\circ}$03’27” W). This receiver was a test for the facilities. Given its location inside the city, it was not suitable for radio solar observations due to its intense background noise. Nevertheless, it took data from February 22nd, 2012 until March 18th, 2015. CALLISTO NA-18 in Pucusana district (Punta Lobos) at the CONIDA scientific site (-12$^{\circ}$30’18” S, -77$^{\circ}$47’56” W), was installed later in 2014. This receiver operated between October 2014 and August 2016 providing data to the e-CALLISTO network. However, after 2016 it was no longer maintained. There are plans to get the instruments back into operation.

\begin{figure}
\centerline{ \includegraphics*[width=0.9\textwidth]{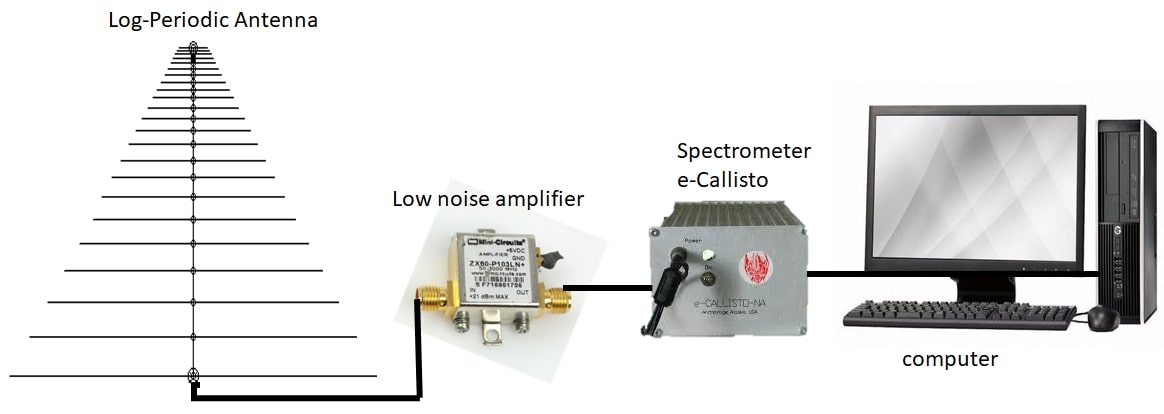}}
  \caption{e-CALLISTO station schematics, including LPDA antenna and spectrometer.}
  \label{fig_station}
\end{figure}

Each station was implemented with an LPDA antenna with 23 elements in a fixed position covering the 70-1000 MHz range. The front-end electronics comprised a low noise amplifier ZX60-33LN-S+ from Mini-Circuits, connected through a coaxial wire to the e-CALLISTO spectrometer, which was built in Anchorage, AK, USA, by W. Reeve. Data were transferred via a RS-232 cable to a computer. The complete setup is shown in Figure \ref{fig_station}.

The CALLISTO spectrometer is a programmable heterodyne receiver designed by C. Monstein (\cite{Benz2005143}, \cite{CALLISTOweb}), that operates in the 45-870 MHz bandwidth and can continuously observe the solar radio spectrum. It can record up to 400 frequencies per spectrum. For 200 channels, the time resolution is 0.25 s and the integration time 1 ms. Each channel has 300 kHz bandwidth. The maximum dynamic range is 40 dB.

It should be noticed that the antenna and spectrometer frequency coverage are slightly different. They superpose between 70 to 870 MHz, while the 45 to 70 MHz range can be recorded by the spectrometer, but is outside the antenna range. On the contrary, from 870 to 1000 MHz the spectrometer cannot register any signal that could be observed by the antenna. 

\subsection{Radio Frequency Interference}

To assess the suitability of the sites the Radio Frequency Interference (RFI) was analyzed. The RFI is a background to the antenna measurements, that can be caused by electromagnetic induction and electromagnetic radiation emitted from external sources. In the worst case, this can turn into a total loss of data. Thus we seek to identify radio-quite zones, free of interference, for radio-astronomy. The {\it National Frequency Allocation Plan} (PNAF \cite{PNAF}) summarizes the frequency bands given in the 9 kHz to 300 GHz range to all telecommunication services in Peru. However, there are geographic considerations and in-situ measurements that must be performed. 

The interference measurements were done with a simple dipole. We used a spectral analyzer INSTEK GSP-GW-830 for frequencies from 9 kHz to 3 GHz, with more power from 45 MHz to 870 MHz. RFI data were taken in 2014 in San Isidro at 10:25 (UT-5) on May 20 and in Pucusana at 14:28 (UT-5) on May 15. 

\begin{figure}
\begin{center}
    \includegraphics*[width=0.9\textwidth]{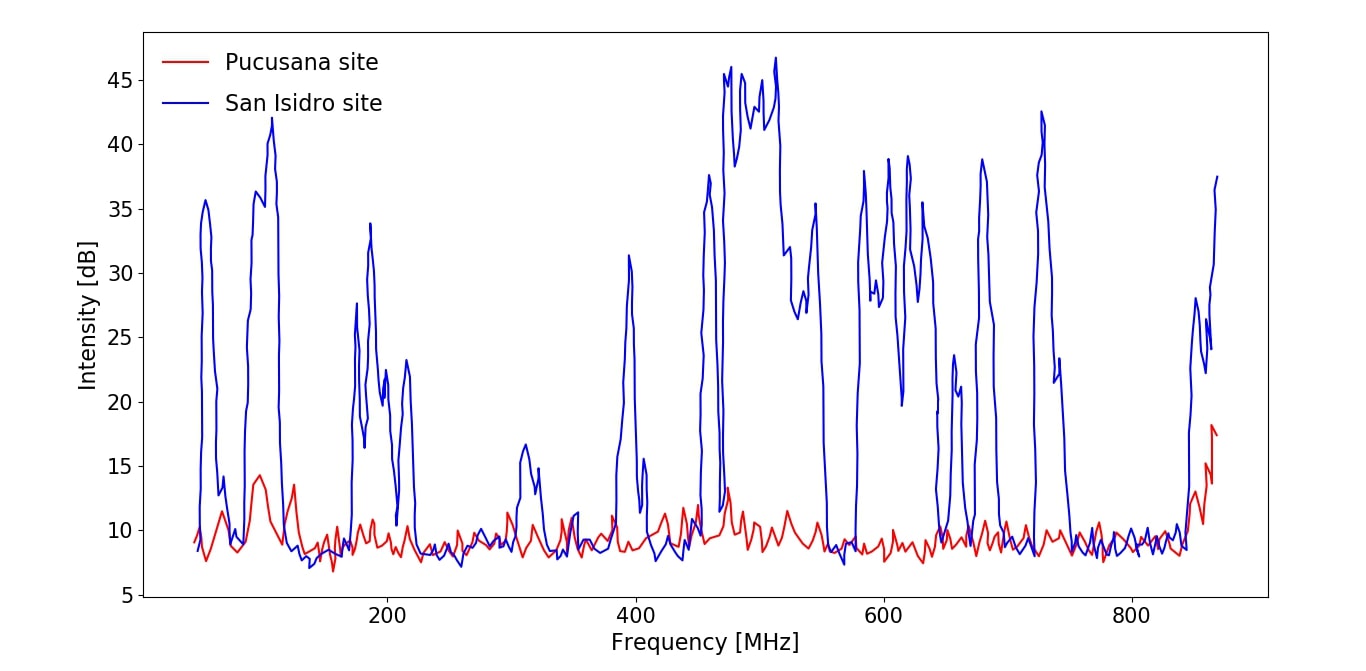}\\
\end{center}
  \caption{Radio Frequency Interference intensity as a function of the frequency, measured with a dipole at San Isidro (blue) and Pucusana (red) sites.}
  \label{fig_RFI}
\end{figure}

In Figure \ref{fig_RFI} we show the RFI intensity as a function of frequency for both sites. The interference in San Isidro is very intense, because it is located inside the city, where there are several telecommunication activities and transmitters. Thus, this receiver could not monitor solar radio phenomena due to its large background during the whole data taking period. Plans to relocate it in the future have been considered. Instead, the Pucusana site has fewer peaks, lower background noise, since it is located in the outskirts of the city and has a natural terrain shielding given by the surrounding hills. Therefore, we consider the Pucusana site to be a good option for SRBs observations. The range of our reported radio bursts observations is between 110 to 440 MHz.

\subsection{Radiation Pattern and Beamwidth}

\begin{figure}[ht]
	   \centering
	   \includegraphics*[width=0.8\textwidth]{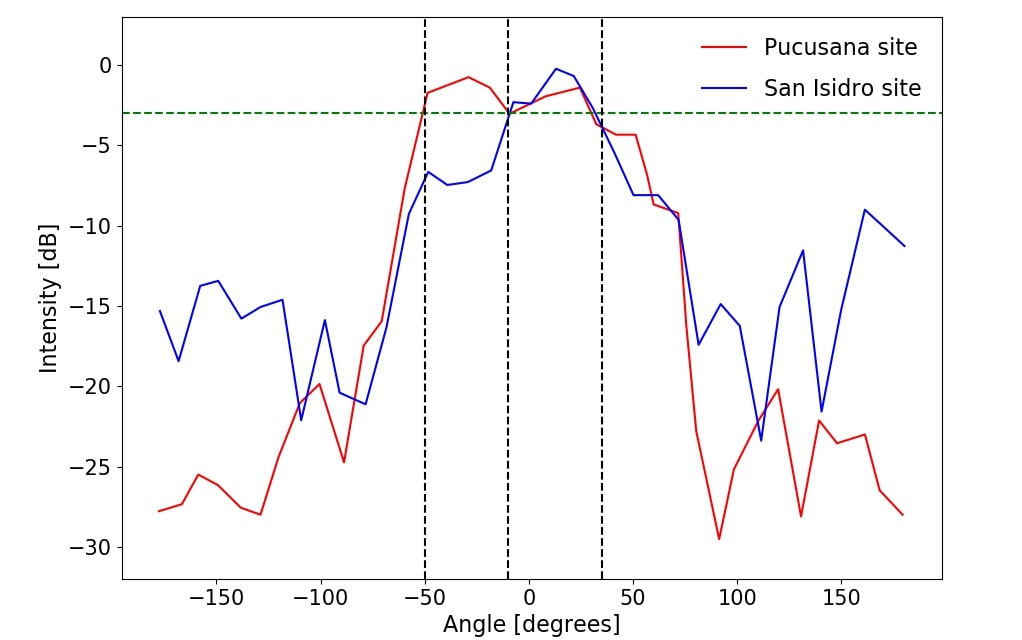}\\
	\caption{LPDA antennas radiation patterns in Cartesian coordinates measured with the spectral analyzer in San Isidro site (blue) and Pucusana site (red). The dashed black lines mark the beam-width at -3 dB (green dashed line).}
\label{fig_radiation_pattern}
\end{figure}

The sensitivity of the LPDA antennas were characterized by determining their radiation pattern and beamwidth. Each antenna was connected to a spectrum analyzer to measure the reception maximum power. The reference signal, centered at 481.76 MHz, was emitted by the repeater antenna of the National Institute of Radio and Television of Peru - IRTP (\cite{IRTP}), located at an altitude of 278 masl in Marcavilca hill (Chorrillos district, Lima). We oriented the antenna horizontally from the largest to the smallest dipole heading for Marcavilca hill located at 8.3 km from the San Isidro site. Once the IRTP signal was found, we oriented the antenna such that it provided the maximum power and from that point, we rotated the antenna a full 360$^{\circ}$ sweep, in 10$^{\circ}$ steps. The same procedure was performed to the antenna at the Pucusana site.

In Figure \ref{fig_radiation_pattern} we show the measurements of the radiation patterns at both sites in Cartesian coordinates. The beam-width, the angle between the points where the signal strength is half of its maximum radiated power (-3 dB), obtained experimentally was 41$^{\circ}$ for the San Isidro site and 43$^{\circ}$ for the Pucusana site. 

\section{Radio Observations with the Peruvian e-CALLISTO station at Pucusana}
\label{Sec_Obs}

The Peruvian e-CALLISTO station at the Pucusana site (i.e. Punta Lobos) took data from October 10th, 2014 until August 3th, 2016, with few missing daily data. These data are stored in the e-CALLISTO network server \cite{e-callisto-Data}, which has the spectra of all instruments of the network. For this study, we searched for radio bursts from the data taken by the CALLISTO NA-18 station (i.e. \textit{CONIDA\_PL} in the server). 

\begin{figure}[h!]
\begin{center}
    \includegraphics*[width=1.00\textwidth]{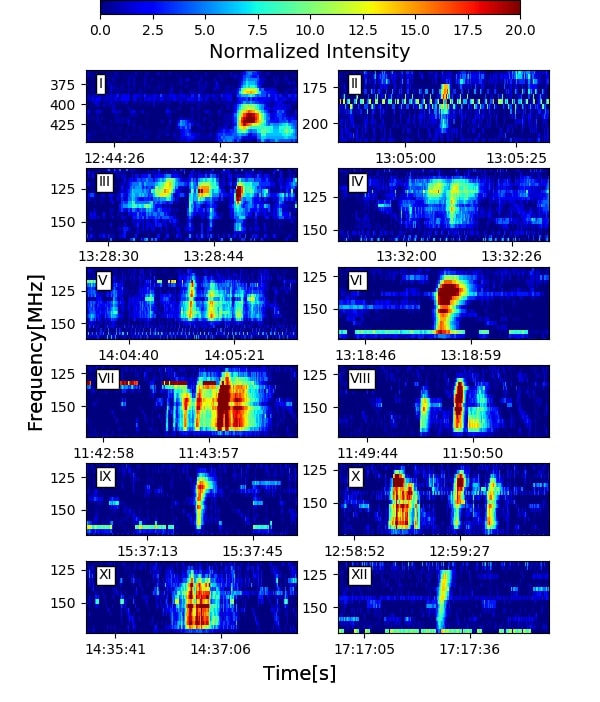}\\
\end{center}
  \caption{Dynamical spectra of type III SRB event candidates taken with CALLISTO NA-18 station at the Pucusana site. Data are centered in the frequency and time ranges of the signal. Specific dates and further parameters are given in Table \ref{Table_burstsdates}.}
\label{FigSpects}
\end{figure}

At the Pucusana site most signals were weak compared to the RFI. After searching the whole dataset for radio bursts, we selected the most significant signals for this analysis, comprising a set of twelve events. These took place on different days between December 2014 and June 2015. In Figure \ref{FigSpects} we show the dynamical spectra of these recorded radio signals, which we consider are type III SRBs with their characteristic shapes, as single or group bursts. The events are ordered using Roman numbers. The first two events are found at a higher frequency range, from 178 to 433 MHz and the remaining events are in the range from 114 to 174 MHz.

The data shown in Figure \ref{FigSpects} have been processed and cleaned, as described in \cite{Chihway2015}, as follows:
\begin{enumerate}
    \item We convert the raw data units into dB by multiplying the data with the conversion factor ($C_f$) that depends on the e-CALLISTO spectrometer: $C_f=2500[
\textrm{mV}]/(255[\textrm{ADU}]\times25.4[\textrm{mV/dB}])$, where ADU stands for Analog Digital Units.
    \item For each frequency, we subtract the median value over time (i.e. 15 min). This step removes the time independent low-level standing-wave pattern.
    \item We apply to the data a multidimensional Gaussian filter, which is implemented as a sequence of one-dimensional Gaussian filters. 
    \item Finally, we normalize the intensity from -5 to 25 dB, obtaining the final spectrum.
\end{enumerate}

It is worth mentioning that the complete spectra for these events found in the database shows, in most cases, a special effect (i.e. a mirrored signal at different frequencies). The real bursts, shown in Figure \ref{FigSpects}, range from 114 to 174 MHz. The mirrored part (i.e. ghost spectrum), at lower frequencies, with an opposite drift rate is not real and thus we do not show it here. This effect is produced by a saturation of the low noise pre-amplifier and/or spectrometer caused by a strong FM-transmitter (88 - 108 MHz according to \cite{PNAF}), which acts as a local oscillator. The semiconductor, when saturated, operates in a non-linear regime and acts like a multiplier, creating the mirror signal at lower frequencies. 

For each dynamical spectra the most intense or isolated radio burst is chosen for a more detailed analysis. The observation date and time, significance cut, frequency range, drift rate and duration are listed in Table \ref{Table_burstsdates}. In addition, for the cases where it was possible, we have calculated a global frequency drift for group bursts. The calculation of these burst parameters is described next. 

\begin{table}[!ht]
\centering
  \begin{tabular}{||c c c c c c c||} 
 \hline
SRB & Date & cut& Frequency Range  &  Drift & Duration& Global Drift \\
 & (UT) & ($\sigma$)& (MHz)  &  (MHz/s) & (s)& (MHz/s)  \\
  \hline 
I & 2014/12/21 - 12:44:39 & 2.5& 411 - 433  & $-84.0\pm28.2$  & $1.39\pm0.02$& s\\
II & 2014/12/21 - 13:05:08 & 1& 178 - 196  & $-41.4\pm$5.4 & 1.32$\pm$0.03& s\\
III & 2015/01/14 - 13:28:47 & 3.5& 122 - 140  & $-15.5\pm$0.5 & 3.12$\pm$0.16& 0.55$\pm$0.35\\
IV & 2015/01/14 - 13:32:10 & 3& 118 - 140  & $-14.0\pm$3.3 &3.77$\pm$0.27& * \\
V & 2015/01/14 - 14:05:02 & 2.5& 131 - 157 & $-35.2\pm$1.8 & 2.11$\pm$0.08& 0.20 $\pm$ 0.14\\
VI & 2015/01/25 - 13:18:54 & 2& 122 - 174 & $-39.2\pm$2.4 & 1.50$\pm$0.02& s \\
VII & 2015/01/26 - 11:44:01 & 4& 127 - 170 & $-23.8\pm$4.6 &  4.48$\pm$1.66 & 0.51$\pm$0.17 \\
VIII & 2015/01/26 - 11:50:37 & 3& 127 - 170 & $-30.2\pm$2.8 &2.91$\pm$0.03& * \\
IX & 2015/01/26 - 15:37:27 & 1.5& 123 - 140 & $-18.8\pm$3.9 & 2.05$\pm$0.03& s\\
X & 2015/02/01 - 12:59:26 & 2& 127 - 149 & $-18.5\pm$1.4 & 1.88$\pm$0.03 & 0.14$\pm$0.02\\
XI & 2015/02/01 - 14:36:38 & 2.5& 127 - 170 & $-25.3\pm $5.8& 2.99$\pm$0.18 & $\dagger$ \\
XII & 2015/06/30 - 17:17:26 & 1& 114 - 170 & $-23.1\pm$1.5  &1.49$\pm$0.02&s\\ 
 \hline
\end{tabular}
\caption{Radio Bursts characteristics corresponding to the numbering shown in Figure \ref{FigSpects}. Note: for the global drift s stands for single (isolated) bursts (i.e. no global drift), * represents group bursts with a weak second signal where no global drift could be calculated and $\dagger$ corresponds to group bursts without a global drift (i.e. constant frequency).}
    \label{Table_burstsdates}
\end{table}

\begin{figure}[ht]
\centering

\includegraphics*[width=0.9\textwidth]{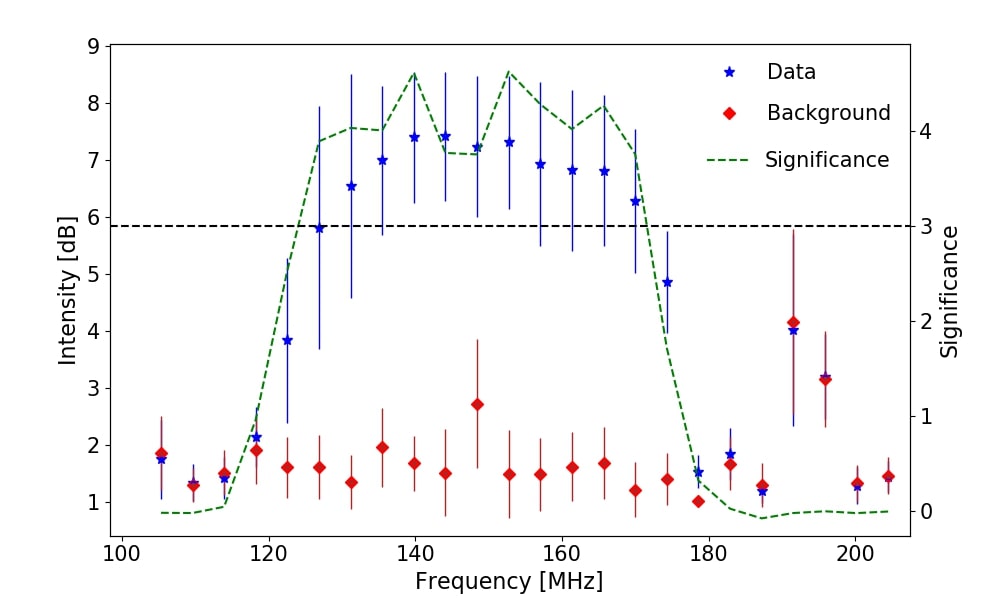}
\hfill
\caption{Spectrum comparison between data (blue starts) and background (red diamonds) for SBR VIII. The background is obtained from a similar measurement from one day before without signal. The green line shows the significance (right axis). The black line is the significance cut, 3 $\sigma$ in this case.}
\label{fig_data_bckg}
 \end{figure}

The burst frequency range is defined where the signal is above the background given a significance cut. We compare the measured spectra with the mean intensity in a 15-minute interval from one day before without signal (i.e. background ($b$)), as seen in Figure \ref{fig_data_bckg} for SRB VIII. We calculate for each frequency the average intensity ($n$) over the time interval of the burst (i.e. $\approx$ 2 s). Using the significance ($S=\frac{n-b}{\sqrt{b}}$), we apply a cut at different $\sigma$s to select the frequency signal range to calculate the drift rate, as described later. The significance cut depends on the strength of the signal and has a lower limit of 1$\sigma$, with an average value of 2.4$\sigma$ for all events. 

\subsection{Type III Solar Radio Bursts Parameters}

The recorded radio spectrum is used to study the characteristics of the type III SRBs. A feature of SRBs is their frequency drift rate $df/dt$ (\cite{Ratcliffe_2014}), which is an indication of fast electron beams moving in the solar corona. The drift rate is the positive or negative displacement of the frequency representing the burst peak flux in a time interval (\cite{Reid_2014}).

The drift rate is given by:
\begin{equation}
    \label{eq_drift}
  D=\frac{df}{dt}=\frac{f_{f}-f_{i}}{t_{f}-t_{i}}
  \end{equation}
where $f_{i}$ and $f_{f}$ are the frequencies corresponding to the peak flux at the starting ($t_{i}$) and end ($t_{f}$) times of the burst, respectively. 

Drift rates have been studied in different frequency ranges. In \cite{Morosan_2018} drift rates vary from -1 MHz/s at 20 MHz to -20 MHz/s at 100 MHz. In \cite{Zhang2018}, they found for the 10–80 MHz range an average drift rate of -7 MHz/s, while in \cite{Meszarosova2008} for a broader frequency study from 950 to 2500 MHz, the estimated drift rate was $\pm$ 500 MHz/s for narrow bursts with bandwidth of about 100 MHz. 

\begin{figure}[ht]
\centering

\includegraphics*[width=0.9\textwidth]{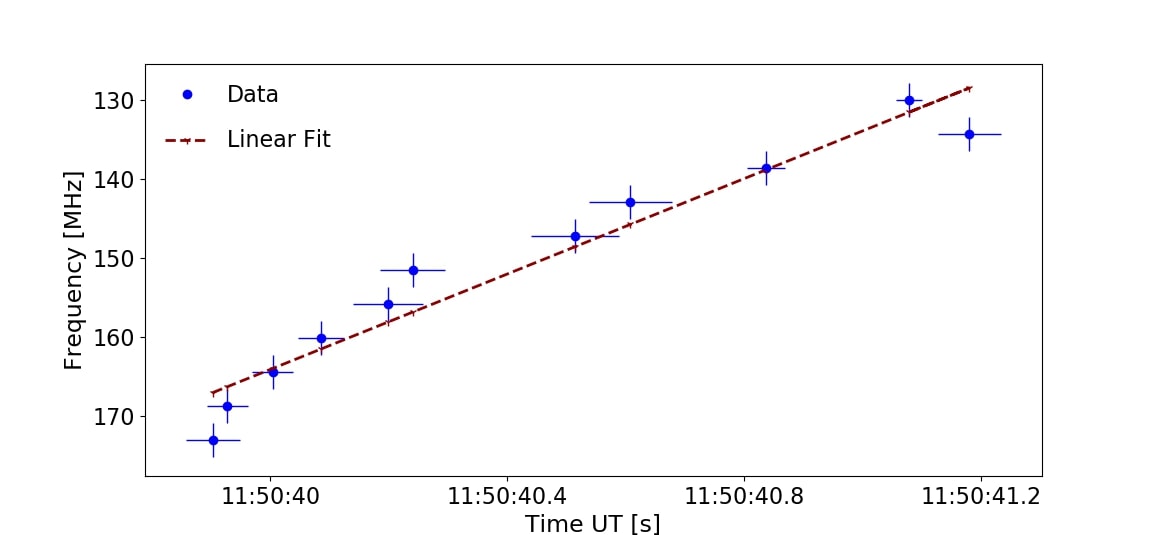}

\caption{Frequencies of the peak flux for each LC and associated times for SBR VIII. The red dashed line represents the linear regression ($\chi^2_{red}$ = 13.8), which in this case shows a negative drift rate.}
\label{fig_linear_fit}
 \end{figure}

To calculate $D$ we follow this procedure. First, we create the light curves of each measured channel (frequency) of the spectrometer, excluding the frequencies with an intensity below a given significance (between 1 and 4 depending on the strength of the signal) as discussed before. 
For each light curve, we fit a Gaussian profile around the peak flux. From the fit we take the time corresponding to the center of the Gaussian, which together with the specific frequency form an ordered pair. This is drawn in a scattered plot as seen in Figure \ref{fig_linear_fit} for SRB VIII, as an example. We observe a linear dependency representing the frequency drift rate, which is estimated from a linear regression of these points.

To obtain the observed duration of the bursts, we get the full width at half maximum (FWHM) of each light curve peak Gaussian fit. Then, the duration is taken as the average of these FWHMs (\cite{Zhang2019}) from all frequencies profiles for each SRB.

\begin{figure}[ht]
\centering

\includegraphics*[width=0.9\textwidth]{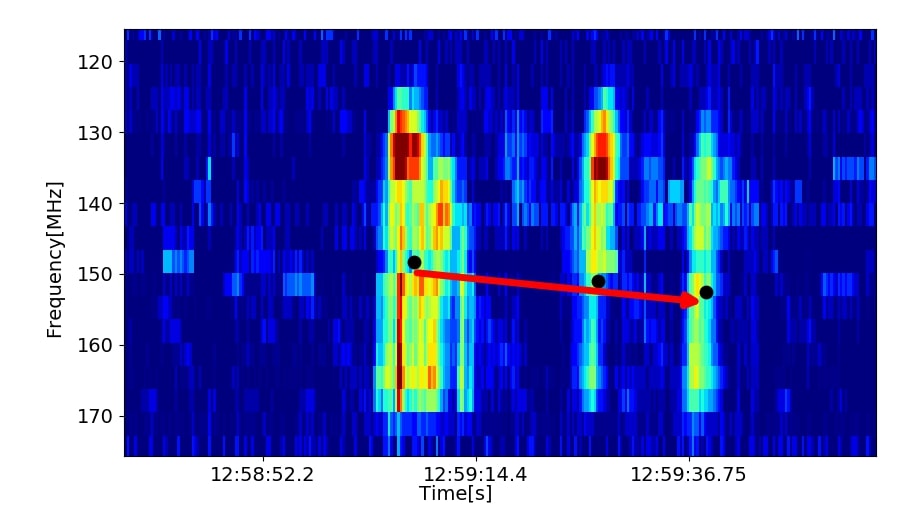}

\caption{Dynamical spectra of SBR X showing a positive global frequency drift marked with a characteristic red arrow. The black points correspond to the central frequencies for each burst.}
\label{fig_GFD}
 \end{figure}

Moreover, we calculated the global frequency drift for solar events with a group of consecutive bursts, defined in \cite{Baolin2008}, as the variation of the central burst frequency during the time of the event, called drifting pulsating structure \cite{Barta2008}. In order to choose the central frequency of each burst, we follow the same procedure described before, applying a  different significance cut to each burst in the group, depending on the signal intensity. In Figure \ref{fig_GFD} we show for SRB VII, as an example, the fitted global frequency drift  indicated with a characteristic arrow. Some studies \cite{Barta2008} suggest that these global drifts occur because the plasmoids, semi-closed magnetic field structures, move in the stratified solar atmosphere.

All observed SRBs have negative drift rates. The first SRB shown in Figure \ref{FigSpects} appears in a higher frequency range from 411 to 433 MHz with drift rate D = -84.0 $\pm$ 28.2 MHz/s. The second SRB is between 178 - 196 MHz and has a drift of -41.4 $\pm$ 5.4 MHz/s. All other observed SRBs are between 114 to 174 MHz and their  calculated drifts have a mean of -25.8 $\pm$ 3.72 MHz/s. Negative drift rates are standard for type III SRBs. This can be attributed to electron beams moving away from the solar surface up to the high corona along open field lines, from regions of high to low densities (\cite{Robinson2000, Baolin2019}). 
The duration of the higher frequency range burst was 1.39 $\pm$ 0.02 s, for the middle frequency range it was 1.32 $\pm$ 0.03 s, while for the lower frequencies it was 2.6 $\pm$ 0.25 s. The mean bandwidth was 28 MHz for the first range, 18 MHz for the second range and 35 MHz for the third range. Individual burst values are given in Table \ref{Table_burstsdates}.

In addition, we calculated a characteristic property of plasmoids, the global frequency drift, for group bursts, excluding isolated events and weak signals, listed in Table \ref{Table_burstsdates}. We found four group bursts, out of the twelve SRB events, with a positive drift with mean 0.4 $\pm$ 0.1 MHz/s. These low values indicate that the drift varies slowly.

\section{Conclusions}
\label{Sec_conclusions}

Two e-CALLISTO stations have been installed and commissioned in Lima, Peru. The main advantage of installing solar monitors in Peru is its strategic location near the Equator, making it possible to observe evenly the Sun throughout the year. In addition, during the data taking period of the analyzed receiver, the Peruvian e-CALLISTO detector was unique in its time-zone coverage. 

We have calibrated the antennas measuring their radiation pattern and beam-width. We have also analyzed the ambient background noise, showing that the San Isidro station had an intense RFI, not suitable for identifying SRBs, since the station was inside the city. This station was installed earlier in 2012 for tests purposes. However, the Pucusana station, installed in 2014, had a lower background due to the natural terrain shielding and distance to the city.

Once the system was commissioned, we took data looking for SRBs candidates between October 10th, 2014 and August 3th, 2016 at the Pucusana site. We have shown that this Peruvian e-CALLISTO station was able to observe type III SRB events in the metric and decimetric bands. Nevertheless, only twelve radio bursts have been identified with a significance larger than 1 $\sigma$ in the whole dataset, since most possible signals were weak compared to the RFI. The most common type III SRB had been characterized with the following values: they are located in the frequency range between 114 to 174 MHz, they have a negative drift rate of -25.8 $\pm$ 3.7 MHz/s, a duration of 2.6 $\pm$ 0.3 s and 35 MHz bandwidth. For some events showing group bursts it was possible to calculate a global frequency drift with mean 0.4 $\pm$ 0.1 MHz/s, characterizing the plasmoids.

\normalem
\begin{acknowledgements}

J. R. appreciates the support from the Peruvian National Council for Science, Technology and Technological Innovation scholarship under Grant 23-2015-FONDECyT and thanks Fernando Valle's help in the installation of CALLISTO. J. B. thanks the Dirección de Gestión de la Investigación (DGI - PUCP) for funding under Grant No. DGI-2019-3-0044.

We thank the Air Force electronic laboratory SELEC for the tests done at their facilities and the Institute for Data Science FHNW Brugg/Windisch, Switzerland for providing data of the e-CALLISTO network. We would also wish to thank C. Consolandi, F. Valle and S. della Torre for reading the manuscript and their useful suggestions. We specially thank C. Monstein for his help and suggestions in the analysis of the e-CALLISTO data and also for reading the manuscript.

\end{acknowledgements}
  
\bibliographystyle{raa}
\bibliography{ms2020-0444.bib}

\begin{thebibliography}{35}
\providecommand\natexlab[1]{#1}
\providecommand\JournalTitle[1]{#1}

\bibitem[Ali {et~al.}(2016)]{Ali2016}
Ali, M., Sabri, S., Hamidi, Z., {et~al.} 2016, ICIMSA 2016 - 3rd International
  Conference on Industrial Engineering, Management Science and Applications

\bibitem[Ansor {et~al.}(2019)]{Ansor2019}
Ansor, N., Hamidi, Z., \& Shariff, N. 2019, Journal of Physics: Conference
  Series, 1349

\bibitem[Axisa(1974)]{Axisa_1974}
Axisa, F. 1974, Solar Physics, 35, 207

\bibitem[Benz {et~al.}(2005)]{Benz2005143}
Benz, A., Monstein, C., \& Meyer, H. 2005, Solar Physics, 226, 143

\bibitem[Benz {et~al.}(2009)]{Benz2009277}
Benz, A., Monstein, C., Meyer, H., {et~al.} 2009, Earth, Moon and Planets, 104,
  277

\bibitem[Bárta {et~al.}(2008)]{Barta2008}
Bárta, M., Karlický, M., \& Žemlicka, R. 2008, Solar Physics, 253, 173–189

\bibitem[Chang {et~al.}(2015)]{Chihway2015}
Chang, C., Monstein, C., Refregier, A., {et~al.} 2015, Astronomy Society of the
  Pacific, 127, 957

\bibitem[{FHNW Windisch}(2020)]{e-callisto-Data}
{FHNW Windisch}. 2020, Observations of generation AOS, Argos, Phoenix-3,
  Phoenix-4 and e-Callisto, accessed: 2020-11-13

\bibitem[Ginzburg \& Zheleznyakov(1958)]{Guinzburg_1958}
Ginzburg, V.~L., \& Zheleznyakov, V.~V. 1958, Soviet Astron. J., 2, 653

\bibitem[Hamidi {et~al.}(2019)]{Hamidi2019}
Hamidi, Z., Ramli, N., \& Shariff, N. 2019, Journal of Physics: Conference
  Series, 1152

\bibitem[Hamidi {et~al.}(2016)]{Hamidi2016}
Hamidi, Z., Zainol, N., Ali, M., {et~al.} 2016, ICIMSA 2016 - 2016 3rd
  International Conference on Industrial Engineering, Management Science and
  Applications

\bibitem[Kasahara {et~al.}(2001)]{kasahara_2001}
Kasahara, Y., Matsumoto, H., \& Kojima, H. 2001, Radio Science, 36, 1701

\bibitem[Labrum(1971)]{Labrum}
Labrum, N.~R. 1971, Australian Journal of Physics, 24, 193

\bibitem[Ma {et~al.}(2012)]{Ma_2012}
Ma, Y., Xie, R., Zheng, X., Wang, M., \& Yi-hua, Y. 2012, Chinese Astronomy and
  Astrophysics, 36, 175

\bibitem[MacAlester \& William(2014)]{McAlester2014}
MacAlester, M.~H., \& William, M. 2014, American Geophysical Union

\bibitem[McCauley(2017)]{Patrick}
McCauley, P.~I. 2017, Astrophysical Journal, 325

\bibitem[Melrose(1980)]{Melrose_1980}
Melrose, D.~B. 1980, Plasma astrophysics: nonthermal processes in diffuse
  magnetized plasmas, Vol.~2 (New York, Gordon and Breach Science Publishers)

\bibitem[{Ministerio de Transporte y Comunicaciones - Peru}(2008)]{PNAF}
{Ministerio de Transporte y Comunicaciones - Peru}. 2008, Plan Nacional de
  Atribución de Frecuencias {PNAF}, accessed: 2020-04-29

\bibitem[{Ministerio de Transporte y Comunicaciones - Peru}(2020)]{IRTP}
---. 2020, {Instituto Nacional de Radio y Televisión del Peru}, accessed:
  2020-03-09

\bibitem[Monstein(2020)]{CALLISTOweb}
Monstein, C. 2020, {e-CALLISTO} International Network of Solar Radio
  Spectrometers, a Space Weather Instrument Array, accessed: 2020-03-09

\bibitem[Morosan \& Gallagher(2018)]{Morosan_2018}
Morosan, D.~E., \& Gallagher, P.~T. 2018, Planetary Radio Emissions VIII

\bibitem[Mészárosová {et~al.}(2008)]{Meszarosova2008}
Mészárosová, H., Karlický, M., Sawant, H.~S., {et~al.} 2008, Astronomy and
  Astrophysics, 484, 529

\bibitem[Prasert {et~al.}(2019)]{Prasert2019}
Prasert, N., Phakam, A., Asanok, K., {et~al.} 2019, Journal of Physics:
  Conference Series, 1380

\bibitem[Ramli {et~al.}(2015)]{Ramli2015123}
Ramli, N., Hamidi, Z., Abidin, Z., \& Shahar, S. 2015, International Conference
  on Space Science and Communication, IconSpace, 123

\bibitem[Ratcliffe {et~al.}(2014)]{Ratcliffe_2014}
Ratcliffe, H., Kontar, E.~P., \& Reid, H. A.~S. 2014, Astronomy \&
  Astrophysics, 572, A111

\bibitem[Reid \& Ratcliffe(2014)]{Reid_2014}
Reid, H. A.~S., \& Ratcliffe, H. 2014, Research in Astronomy and Astrophysics,
  14, 773–804

\bibitem[Robinson \& Benz(2000)]{Robinson2000}
Robinson, P., \& Benz, A. 2000, Solar Physics, 194, 345

\bibitem[Suzuki \& Dulk(1985)]{Suzuki1985}
Suzuki, S., \& Dulk, G.~A. 1985, Solar Radiophysics, 289

\bibitem[Tan(2008)]{Baolin2008}
Tan, B.~L. 2008, Solar Physics, 253, 117–131

\bibitem[Tan {et~al.}(2019)]{Baolin2019}
Tan, B.~L., Naihwa, C., Ya-Hui, Y., {et~al.} 2019, The Astrophysical Journal,
  885

\bibitem[White(2007)]{White}
White, S.~M. 2007, Asian Journal of Physics, 16, 189

\bibitem[Yue {et~al.}(2018)]{Xinan2018}
Yue, X., Schreiner, W.~S., Kuo, Y., {et~al.} 2018, Extreme Events in Geospace,
  541

\bibitem[Zavvari {et~al.}(2016)]{Zavvari2016185}
Zavvari, A., Islam, M., Anwar, R., {et~al.} 2016, Experimental Astronomy, 41,
  185

\bibitem[Zhang {et~al.}(2018)]{Zhang2018}
Zhang, P., Wang, C., \& Ye, L. 2018, Astronomy and Astrophysics, 618

\bibitem[Zhang {et~al.}(2019)]{Zhang2019}
Zhang, P., Yu, S., Kontar, E., \& Wang, C. 2019, Astrophysical Journal, 885

\end{thebibliography}

\end{document}